\documentclass[conference]{IEEEtran}
\IEEEoverridecommandlockouts
\usepackage{cite}
\usepackage{amsmath,amssymb,amsfonts}
\usepackage{algorithmic}
\usepackage{graphicx}
\usepackage{textcomp}
\usepackage{xcolor}
\usepackage{url}
\usepackage{booktabs}
\usepackage{multirow}
\usepackage{bm}
\usepackage[normalem]{ulem}
\useunder{\uline}{\ul}{}
\usepackage{hyperref}
\usepackage{adjustbox}
\usepackage{threeparttable}
\usepackage{makecell}
\usepackage{xspace}
\usepackage[table,xcdraw]{xcolor}

\newcommand{\dataseturl}{uniqueshipdata.org}
\newcommand{\datasetlink}{\href{https://www.uniqueshipdata.org}{uniqueshipdata.org}\xspace}

\def\BibTeX{{\rm B\kern-.05em{\sc i\kern-.025em b}\kern-.08em
    T\kern-.1667em\lower.7ex\hbox{E}\kern-.125emX}}

\usepackage{geometry}
\begin{document}

\title{UniqueShip: Mitigating Data Leakage in Acoustic Ship Classification Benchmark Datasets \vspace{-4pt} \\
}

\author{
  \IEEEauthorblockN{
    Connor Hashemi$^{*\dagger}$\thanks{\noindent $^*$Equal contribution. \hspace{2mm} $^{\dagger}$Corresponding author.},
    Trevor Stout$^*$,
    Anthony Hoogs,
    Jason Parham
  }
  \IEEEauthorblockA{
    \textit{Kitware, Inc.}, 1712 Route 9, Suite 300, Clifton Park, NY 12065, USA \\
    \{connor.hashemi, trevor.stout, anthony.hoogs, jason.parham\}@kitware.com
  }
}


\IEEEaftertitletext{\vspace{-2\baselineskip}}

\maketitle

\begin{abstract}
Underwater Acoustic Target Recognition (UATR) of ships is well-suited for machine learning, yet its progress is hindered by the lack of large, diverse, and publicly available labeled datasets. 
In this work, we introduce \textit{UniqueShip}, a machine learning-ready benchmark dataset for UATR applications sourced from the open Ocean Networks Canada (ONC) repository.
Unlike previous datasets, we explicitly control for ``data leakage'' between the training and evaluation sets to ensure more reliable and generalizable model evaluation that does not encourage the model to memorize individual ships.
We demonstrate that typical, random data partitioning in two prominent UATR datasets leads to falsely optimistic test performance, increasing accuracy by 10--48 percentage points compared to our more careful partitioning.
Ablations on UniqueShip further show that doubling the number of unique vessels improves accuracy by 2.4–2.6 percentage points, while doubling total audio duration improves only by 0.8–1.3 points, indicating that vessel diversity should drive dataset curation more than total hours.
We provide baselines with convolutional and transformer backbones, and analyze how ship metadata correlates with classification performance, finding that individual vessel characteristics predict classification difficulty far better than distance to the hydrophone alone.
Overall, UniqueShip contains 2,460 hours of ship-radiated audio from 4,218 unique vessels (3,437 hours including background). 
We publish the dataset, code, and easy-to-download splits at \datasetlink to foster further UATR research.

\end{abstract}



\begin{figure}
  \centering
   \includegraphics[width=1\linewidth]{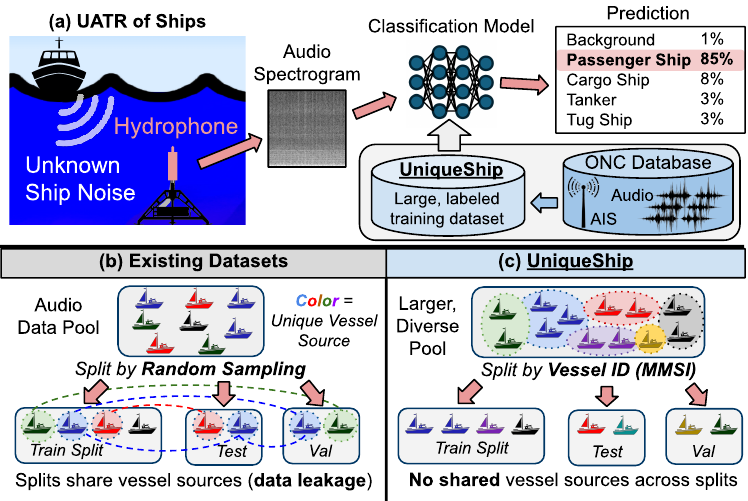}
   \caption{We introduce UniqueShip, a large, labeled, and publicly-available Underwater Acoustic Target Recognition (UATR) dataset sourced from the Ocean Networks Canada (ONC) repository. (a) Our dataset enables robust classification model training. Unlike (b) existing datasets that suffer from data leakage due to random sampling, (c) UniqueShip provides a larger, more diverse pool of distinct recordings and strictly partitions data by Vessel ID (MMSI) to ensure no shared vessels exist across train, test, and validation splits.}
   \vspace{-1.5em}
   \label{fig:intro}
\end{figure}


\section{Introduction}

Classifying vessels using underwater acoustics is crucial for defense, soundscape analysis, and environmental monitoring.  It is hard, however, to obtain and curate passive acoustic data, so labeled datasets for Underwater Acoustic Target Recognition (UATR) are often much smaller than other machine learning (ML) domains. Existing publicly available datasets such as DeepShip \cite{irfan2021deepship}, VTUAD \cite{domingos2022vtuad}, and Oceanship \cite{li2024oceanship} are limited by insufficiently distinct and varied recordings. This lack of data and sample diversity makes deep learning for UATR poorly-generalizable and difficult, especially when trying to use large state-of-the-art models.

In addition to poor data quantity, UATR datasets often contain significant biases. These datasets are composed of continuous acoustic records that exhibit strong temporal and spectral self-similarity, with persistent signatures for individual vessels. Na\"ively partitioning hydrophone recordings into train and test sets using conventional methods (e.g., random sampling of data files or shorter record segments) introduces ``\textit{data leakage}'' into model training and yields falsely optimistic testing results (i.e., overfitting) because the test data is insufficiently distinct from the train partition. This effect may arise in the classification task, for instance, when a model is distracted by the specific acoustic characteristics of an individual ship, hindering the model's ability to differentiate broader ship categories in deployment. We find that when train/test splits are provided in a UATR dataset, they typically include this data leakage, thereby compromising the reported classification accuracy. Therefore, the development of generalizable UATR models \cite{domingos2025improving} is currently constrained by the limited size of available datasets and their suboptimal data partitioning.



To mitigate these limitations, we introduce \textbf{UniqueShip}, a publicly-available benchmark UATR dataset designed specifically to reduce data leakage. See Figure \ref{fig:intro}. Sourced from Ocean Networks Canada (ONC) \cite{heesemann2014ocean}, UniqueShip is the largest open-source machine learning-ready dataset for classifying ship-radiated noise, featuring more than 2,460 hours of ship-radiated audio from 4,218 unique vessels and 977 hours of background audio (see Table~\ref{tab:dataset_comparison}). Rather than sampling splits at random, we partition by vessel identity (MMSI) and group ambient recordings by day, so no vessel or contiguous stretch of background spans multiple splits. Applying these rules to DeepShip \cite{irfan2021deepship} and VTUAD \cite{domingos2022vtuad} exposes a large drop in classification accuracy due to data leakage, and repeating the comparison on UniqueShip confirms the effect. Ablations over vessel count and audio duration establish that diversity is two to three times as valuable per doubling as quantity, and we provide baselines across vision backbones. A metadata analysis shows that classification difficulty is largely encoded in vessel identity and physical vessel metadata, with distance to the hydrophone adding little on its own within the inclusion radius.

\begin{figure}
  \centering
   \includegraphics[width=1\linewidth]{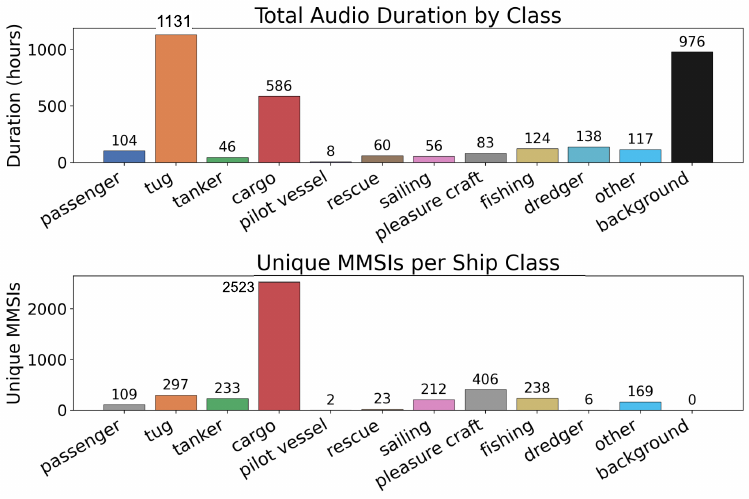}
   \caption{Class distribution of UniqueShip; 11 vessel classes, 1 background class, 4,218 unique vessels, 2,779 days, and nearly 2.5 million 5-sec. samples.}
   \vspace{-1.5em}
   \label{fig:data_volume}
\end{figure}

\section{Open-Source Ship-Radiated Noise Datasets}

\begin{table*}
\centering
\caption{Comparison of Public Datasets Sourced from ONC \cite{heesemann2014ocean, ONC_2556_2022, ONC_2556_2021, ONC_2556_2018, ONC_2523_2023, ONC_2523_2017b, ONC_2523_2017a, ONC_2523_2016}}
\label{tab:dataset_comparison}
\begin{adjustbox}{max width=\textwidth}
\begin{threeparttable}
\begin{tabular}{@{} l c c cc cc cc cc cc cc @{}}
\toprule
\multirow{2}{*}{\textbf{Dataset}} & \multirow{2}{*}{\textbf{Classes}} & \textbf{Inc. / Exc.} & \multicolumn{2}{c}{\textbf{Passenger Ship}} & \multicolumn{2}{c}{\textbf{Cargo}} & \multicolumn{2}{c}{\textbf{Tug}} & \multicolumn{2}{c}{\textbf{Tanker}} & \multicolumn{2}{c}{\textbf{Background}} & \multicolumn{2}{c}{\textbf{Total}} \\
\cmidrule(lr){4-5} \cmidrule(lr){6-7} \cmidrule(lr){8-9} \cmidrule(lr){10-11} \cmidrule(lr){12-13} \cmidrule(lr){14-15}
& & \textbf{(km)} & \textbf{MMSIs} & \textbf{Dur.} & \textbf{MMSIs} & \textbf{Dur.} & \textbf{MMSIs} & \textbf{Dur.} & \textbf{MMSIs} & \textbf{Dur.} & \textbf{MMSIs} & \textbf{Dur.} & \textbf{MMSIs} & \textbf{Dur.} \\
\midrule

DeepShip \cite{irfan2021deepship} & 4 & 2 / 4 & 46 & 12h 22m & 69 & 10h 40m & 17 & 11h 17m & 133 & 12h 45m & --- & --- & 265 & 47h 04m \\
VTUAD (2km/4km) \cite{domingos2022vtuad} & 5 & 2 / 4 & 8 & 2h 04m & 60 & 2h 33m & 25 & 2h 24m & 10 & 2h 06m & --- & 2h 36m & 104 & 11h 43m \\
VTUAD (3km/5km) \cite{domingos2022vtuad} & 5 & 3 / 5 & 8 & 4h 04m & 58 & 4h 24m & 34 & 4h 09m & 7 & 3h 41m & --- & 4h 34m & 108 & 20h 52m \\
VTUAD (4km/6km) \cite{domingos2022vtuad} & 5 & 4 / 6 & 9 & 3h 17m & 47 & 3h 37m & 31 & 3h 21m & 5 & 2h 52m & --- & 3h 38m & 93 & 16h 45m \\
Oceanship \cite{li2024oceanship} & 15 & 1.5 / 3.5 & 60 & 8h 49m & 221 & 30h 56m & 193 & 45h 53m & 19 & 0h 10m & --- & --- & 892 & 121h 10m \\
\addlinespace
\hline \hline
\addlinespace
UniqueShip (5 Class - 5h Each) & 5 & 2 / 8 & 112 & 5h 00m & 2526 & 5h 00m & 304 & 5h 00m & 233 & 5h 00m & --- & 5h 00m & 3175 & 25h 00m \\
UniqueShip (12 Class - 5h Each) & 12 & 2 / 8 & 110 & 5h 00m & 2522 & 5h 00m & 297 & 5h 00m & 233 & 5h 00m & --- & 5h 00m & 4218 & 60h 00m \\
UniqueShip (5 Class - Balanced) & 5 & 2 / 8 & 112 & 42h 40m & 2526 & 42h 40m & 304 & 42h 40m & 233 & 42h 40m & --- & 42h 40m & 3175 & 213h 18m \\
UniqueShip (5 Class - All Data) & 5 & 2 / 8 & 112 & 104h 46m & 2526 & 586h 35m & 304 & 1131h 36m & 233 & 46h 15m & --- & 935h 04m & 3175 & 2804h 16m \\
UniqueShip (All) & 12 & 2 / 8 & 109 & 107h 53m & 2523 & 586h 42m & 297 & 1140h 07m & 233 & 46h 32m & --- & 976h 53m & 4218 & 3436h 39m \\

\bottomrule
\end{tabular}
\begin{tablenotes}
    \small
    \item \textit{Note:} Inc. / Exc. = Inclusion and Exclusion radii, Dur. = Duration. Oceanship \cite{li2024oceanship} MMSI statistics are only from the FG split. Vessels keep only their majority-label samples, falling back to the most frequent included label when the majority is excluded, hence small per-class MMSI shifts across splits.
\end{tablenotes}
\end{threeparttable}
\end{adjustbox}
\vspace{-1.5em}
\end{table*}


\label{sec:onc_datasets}

\textbf{Ocean Networks Canada (ONC) \cite{heesemann2014ocean}}: Ocean Networks Canada (ONC) initiative \cite{heesemann2014ocean} contains a large open-source repository of hydrophone and Automatic Identification System (AIS) data, which can be parsed to extract and classify ship-radiated noise signatures from distinct vessels. This data can be parsed via the Oceans 3.0 Data Portal \cite{oceans30}. Of particular importance is the data collected from hydrophone devices 2556 and 2523 in the Strait of Georgia in the Fraser River Delta region from May 2016 to November 2023 \cite{ONC_2556_2022, ONC_2556_2021, ONC_2556_2018, ONC_2523_2023, ONC_2523_2017b, ONC_2523_2017a, ONC_2523_2016}. These hydrophones are Ocean Sonics icListen AF Hydrophones \cite{oceansonics_iclisten_af} with sample rate of 32kHz and placed at a depth of approximately 100m. Currently, there are three popular datasets which utilize this ONC deployment range: Deepship dataset \cite{irfan2021deepship}, Vessel Type Underwater Acoustic Data (VTUAD) \cite{domingos2022vtuad}, and Oceanship \cite{li2024oceanship}. Smaller works which do not publish their dataset include Yang et al. \cite{yang2019deep}, Tian et al. \cite{tian2021deep}, and Shen et al. \cite{shen2020ship}. However, unlike UniqueShip, none of these available datasets encompass the full range of the deployments nor the full range of metadata available in the AIS data.


\textbf{DeepShip \cite{irfan2021deepship}:} The Deepship dataset \cite{irfan2021deepship} is a popular ship-radiated dataset which contains the ONC deployment data from 02 May 2016 to 04 October 2018 \cite{ONC_2523_2016, ONC_2523_2017a, ONC_2523_2017b}. It collects ship-radiated noise from ships within 2km of the hydrophone and classifies them to 4 ship classes: cargo, passenger, tug, and tanker.




\textbf{Vessel Type Underwater Acoustic Data (VTUAD) \cite{domingos2022vtuad}:} The VTUAD dataset \cite{domingos2022vtuad} contains data captured from 24 June 2017 to 03 November 2017 \cite{ONC_2523_2017a}. In addition to the 4 ship classes in DeepShip, they also collect background data for a total of 5 classes. They define three inclusion/exclusion radii scenarios: 2km/4km, 3km/5km, and 4km/6km. For example, the 4km/6km would record any ship within 4km of the hydrophone as ship-radiated noise and only save background data if no ship is within 6km. The dataset also uses preprocessing to remove signals that are similar to the background and divides the resulting audio into 1-second segments. A train/val/test split is provided which appears to randomly select from the entire corpus of 1-second segments.



\textbf{Oceanship \cite{li2024oceanship}:}
The Oceanship dataset \cite{li2024oceanship} is a ship-radiated dataset based on the ONC repository from July 2020 to February 2021. Unlike previous captures, they include additional vessel labels, going from 4 to 15 different vessel classes.

\section{UniqueShip Dataset and Data Leakage}
We present \textit{UniqueShip}, an open-source dataset based on the ONC repository which emphasizes diversity of unique vessel MMSIs. We utilize similar classification and interval protocols as previous works, but unlike previous datasets we utilize the full range of ONC deployments and eliminate data leakage in the splits. We publish our easy-to-download splits at \datasetlink.

\begin{figure}
  \centering
  \includegraphics[width=1\linewidth]{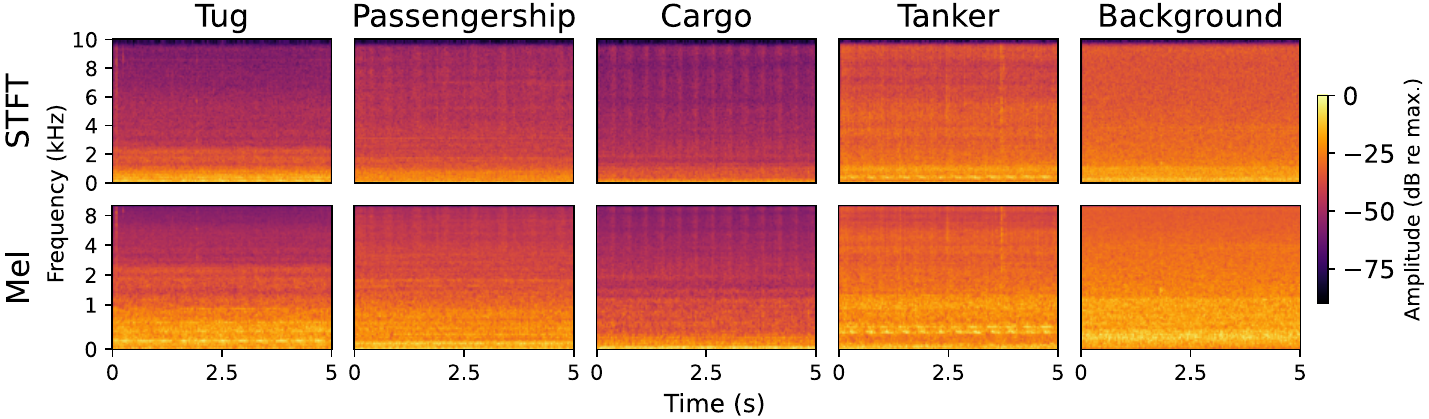}
  \caption{Spectrograms (top) and Mel spectrograms (bottom) from UniqueShip 5 Class subset after preprocessing for input into baseline models (Sec.~\ref{sec:baselines_ablations}).}
  \label{fig:spectrogram_examples}
  \vspace{-1em}
\end{figure}

\subsection{Dataset Description}
\textbf{Source:} As mentioned earlier in Section \ref{sec:onc_datasets}, we combine seven deployments from two icListen AF hydrophones \cite{oceansonics_iclisten_af} from the ONC repository in the Strait of Georgia in the Fraser River Delta region starting from May 2016 to November 2023 \cite{ONC_2556_2022, ONC_2556_2021, ONC_2556_2018, ONC_2523_2023, ONC_2523_2017b, ONC_2523_2017a, ONC_2523_2016}. The hydrophones have a sample rate of 32KHz and are placed approximately 100m below surface. Subsets of these deployments are utilized in the other popular ONC datasets \cite{irfan2021deepship,domingos2022vtuad, li2024oceanship}.

\textbf{Processing:} We built and processed the dataset by modifying the open-source pipeline laid out in previous work \cite{domingos2022vtuad, cesar_lucascesarfdonc_dataset_2026}. We apply a 2km single-vessel inclusion radius and associate all relevant AIS metadata with the collected data. An 8km exclusion radius is used to determine background noise, where an audio clip is characterized as ``background'' if no ships are within 8km radius. This is a larger exclusion radius than Deepship and VTUAD, which results in quieter and less noisy background signals. The audio clips are resampled to 20 kHz and segmented into 5-second samples for best processing in our machine learning models.




\textbf{Available Vessel Metadata:}
We extract labels from the AIS signals, including 17 vessel-specific metadata such as MMSI, distance to hydrophone, speed over ground (SOG), course over ground (COG), ship length and beam size, draught, and navigation status.



\textbf{Overall Statistics}: Figure \ref{fig:data_volume} depicts statistics on the entire UniqueShip dataset, which includes more than 2,460 hours of ship-radiated noise, 977 hours of background audio with no ship present, 4,218 vessels, and 11 ship classes alongside one background class.

\textbf{UniqueShip Splits:}
We construct multiple splits of UniqueShip to facilitate analysis and benchmarks based on specific goals as depicted in the bottom rows of Table \ref{tab:dataset_comparison}. In this paper, we focus on the ``UniqueShip (5 Class - Balanced)'' split, which contains five classes with equal recording duration:  \texttt{background}, \texttt{tug}, \texttt{tanker}, \texttt{cargo}, and \texttt{passenger}. We perform class balancing so that all MMSIs are maintained and have equal representation in the classes. This balanced subset is more than four times larger (by duration) and has twelve times more vessels (by MMSI) than the popular DeepShip benchmark \cite{irfan2021deepship}, and 70\% larger in duration than Oceanship \cite{li2024oceanship}. The subset is further divided into train/test/validation partitions with an approximately 80/10/10\% split with five repetitions. Our repository at \datasetlink includes additional splits and subsets, such as unbalanced splits with all possible data ``UniqueShip (5 Class - All)'', smaller duration splits for expedited or resource-constrained analysis ``UniqueShip (5 Class - 5h Each)'', and splits which include all of the available classes ``UniqueShip (12 Class - 5h Each)''. 



\subsection{Data Leakage in UATR Datasets}

We posit that the training, test, and validation splits in many ship-radiated noise datasets contain shared signals and/or noise which ML models can memorize to increase test performance. In these cases of data leakage, the test set becomes an unreliable source for assessing generalizability to new situations. Existing benchmark datasets \cite{irfan2021deepship,domingos2022vtuad, li2024oceanship} and their users typically create train/test splits by randomly selecting from individual data files, or by segmenting all audio and then randomly separating the shorter segments. Therefore, a single ship, or temporally-adjacent background samples, can appear in both training and testing sets, encouraging researchers and developers to unintentionally overfit their models. This is an emerging topic that is being explored by other works, such as \cite{domingos2025improving}.

\textbf{Minimizing Data Leakage:}
To ensure a reliable evaluation, we propose three often-missed partitioning rules for the benchmark splits. First, samples containing vessel sounds must be grouped by MMSI within the same data split. Second, ambient recordings (i.e., noise) must be grouped by day. Third, the dataset must provide explicit train/test splits or cross-validation folds that enforce these constraints. We expect these rules will significantly minimize data leakage and better reflect real-world inference conditions.

\textbf{Data Leakage Results in DeepShip and VTUAD:}
As evidence of data leakage in existing benchmarks, we apply these rules to the DeepShip \cite{irfan2021deepship} and VTUAD \cite{domingos2022vtuad} datasets using their published metadata to construct new splits. We evaluate five independent 80/10/10 train/val/test splits using the Swin \cite{swin} backbone and our standard baseline settings (see Sec.~\ref{sec:baselines_ablations}). To demonstrate the impact of data leakage, we compare our careful splits against conventional na\"ive random partitioning of the audio files (longer data records for DeepShip and shorter segments for VTUAD), which causes ships to be shared between splits for both datasets. Results are shown in Table \ref{tab:DeepShip_VTUAD_data_leakage} including for the single partition provided with VTUAD (``Published Split" rows). As evident in both DeepShip and VTUAD, the ``No Leakage" rows achieve much worse accuracy and F1 score than their ``Data Leakage'' counterparts, resulting in 13--14 point drop in F1 for DeepShip and a 61 point drop for VTUAD. These drops in F1 score are clear evidence that \textit{data leakage exists in prominent UATR datasets as typically used} and must be carefully avoided.

\begin{table}
    \centering
    \footnotesize
    \setlength{\tabcolsep}{3pt} 
    \caption{Swin model results on prominent benchmark datasets.}
    \begin{tabular}{c l l c c} 
        \toprule
        \textbf{Dataset} & \textbf{Config.} & \textbf{Trans.} & \textbf{Accuracy} & \textbf{F1} \\
        \midrule
        \multirow{4}{*}{\makecell{DeepShip \\ \cite{irfan2021deepship}}} & \multirow{2}{*}{No Leakage} & STFT & $0.583 \pm 0.091$ & $0.548 \pm 0.110$ \\
        & & Mel & $0.588 \pm 0.077$ & $0.563 \pm 0.084$ \\
        \cmidrule(lr){2-5}
        & \multirow{2}{*}{Data Leakage} & STFT & $0.685 \pm 0.030$ & $0.675 \pm 0.031$ \\
        & & Mel & $0.710 \pm 0.029$ & $0.700 \pm 0.031$ \\
        \midrule
        \multirow{6}{*}{\makecell{VTUAD \\ (4km/6km) \\ \cite{domingos2022vtuad}}} & \multirow{2}{*}{No Leakage} & STFT & $0.345 \pm 0.046$ & $0.207 \pm 0.083$ \\
        & & Mel & $0.416 \pm 0.069$ & $0.279 \pm 0.092$ \\
        \cmidrule(lr){2-5}
        & \multirow{2}{*}{Data Leakage} & STFT & $0.821 \pm 0.002$ & $0.821 \pm 0.002$ \\
        & & Mel & $0.888 \pm 0.003$ & $0.888 \pm 0.003$ \\
        \cmidrule(lr){2-5}
        & \multirow{2}{*}{Published Split} & STFT & $0.805 \phantom{{}\pm 0.000}$ & $0.737 \phantom{{}\pm 0.000}$ \\
        & & Mel & $0.893 \phantom{{}\pm 0.000}$ & $0.844 \phantom{{}\pm 0.000}$ \\
        \bottomrule
    \end{tabular}
    \label{tab:DeepShip_VTUAD_data_leakage}
    \vspace{-1.5em}
\end{table}

\section{UATR Baselines and Ablations}
\label{sec:baselines_ablations}

We perform three experiments on the UATR ship classification task: 1) We demonstrate baseline performance on our leak-free dataset split, 2) We intentionally introduce data leakage and record the difference in results, and 3) We evaluate the reliance of classification performance on ship diversity and audio duration. As mentioned in the previous section, these experiments are performed on the ``UniqueShip (5 Class - Balanced)'' dataset to match previous benchmarks \cite{irfan2021deepship, domingos2022vtuad}.

\textbf{Model Training:} To train baseline models, we attach a shallow two-layer classification head to a CNN-based (MobileNetV3-Small \cite{howard2019mobilenetv3}) or transformer-based (SwinV2-Tiny \cite{liu2022swinv2} or ViT-B/16 \cite{dosovitskiy2020ViT}) ImageNet-pretrained backbone. In addition to previous dataset processing, the samples are also filtered to remove the DC component and windowed to remove edge discontinuities. Data are then transformed into spectrograms with 90-dB dynamic range using the Short-Time Fourier Transform (STFT), with and without applying a Mel filter-bank. The STFT method uses 512-sample window Fourier transforms with 50\% overlap, while the Mel method uses 1024-sample windows (with 100\% zero padding) and 75\% overlap, so that the output data size is roughly square and can be easily resized to the input image size of 224x224 (MobileNet and ViT) or 256x256 (Swin) pixels. Figure \ref{fig:spectrogram_examples} illustrates example input data samples. Models are fine-tuned for 30 epochs with the AdamW optimizer and cross entropy loss, 0.01 weight decay factor, and a 1e-6 learning rate reduced by a factor of 5 when the validation loss plateaus.

\textbf{Results with Leakage vs No Leakage:} Baseline ML results are depicted in Table \ref{table:results} when fine-tuning on 1) UniqueShip (``No Leakage") or 2) a na\"ive random sampling (``Data Leakage"). Average values are given across the five folds. For all model settings, average accuracy increases by 0.08--0.21 and F1 increases by 0.09--0.21 with data leakage. In addition to previous dataset experiments (Table \ref{tab:DeepShip_VTUAD_data_leakage}), this is another strong indicator that \textit{data leakage leads to falsely optimistic test results on UATR benchmarks}. Overall, without data leakage, the Swin backbone outperforms the other models tested, and the Mel spectrogram outperforms the simple STFT spectrogram in all cases.

\begin{figure}
  \centering
  \includegraphics[width=1\linewidth]{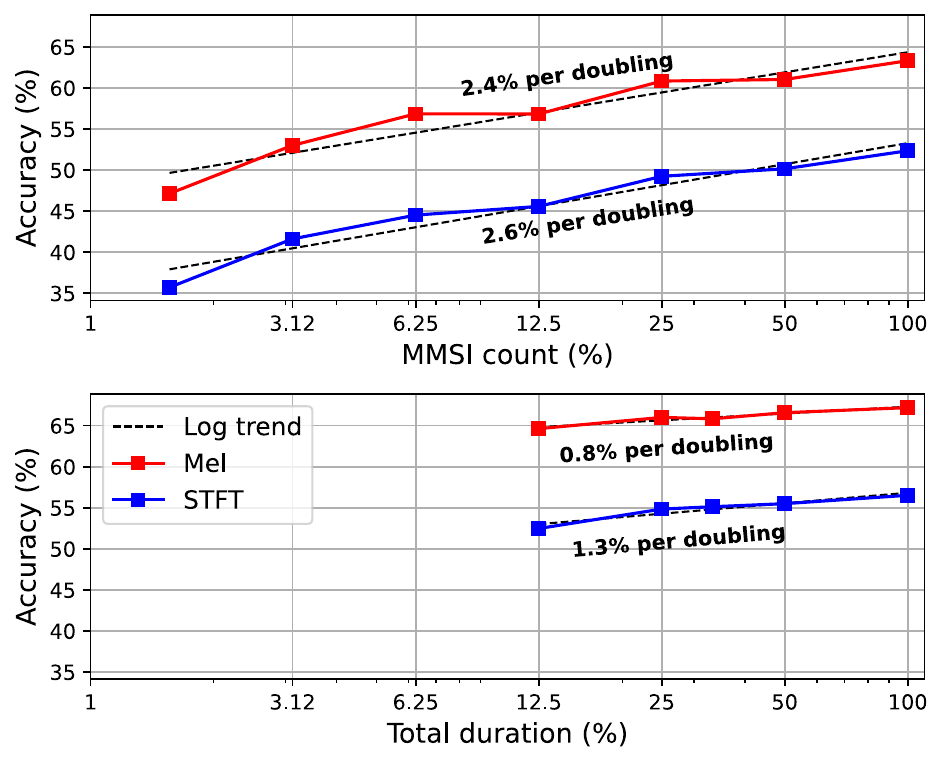}
  \caption{MMSI (top) and duration (bottom) ablation studies on UniqueShip subsets using the Swin backbone. Both vessel count (top) and audio duration (bottom) scale log-linearly with accuracy, but at different rates: 2.4–2.6 points per doubling of unique MMSIs against 0.8–1.3 per doubling of duration. Differing 100\% base durations account for the offset baselines.}
  \label{fig:mmsi_duration_ablation}
  \vspace{-1em}
\end{figure}

\textbf{Results on MMSI and Duration Ablations}: To investigate the performance reliance on dataset statistics, we perform two ablation studies: one in which we hold the audio duration per class constant (5 hours) and vary the amount of available unique MMSIs (ships), and another in which we keep the MMSIs constant (3175 total) but change the audio duration per class with a max of 42 hours per class while holding total training iterations roughly constant. The results are depicted in Fig \ref{fig:mmsi_duration_ablation} with a log scale on the varied parameter. Interestingly, both studies exhibit a log-linear trend, where a doubling of ship count results in 2.4--2.6 percentage point increase in accuracy, and doubling total duration has a weaker effect with 0.8--1.3 increase. This result is evidence that while data diversity and quantity are crucial for UATR models, \textit{vessel diversity has a larger impact}, thereby future datasets should focus on increasing number of unique vessel sources rather than total audio quantity. 


\begin{table}[t]
    \centering
    \footnotesize
    \setlength{\tabcolsep}{4pt}
    \caption{Baseline model test results on UniqueShip subset.}
    \begin{tabular}{c l c c c}
        \toprule
        \textbf{Config.} & \textbf{Backbone} & \textbf{Trans.} & \textbf{Accuracy} & \textbf{F1} \\
        \midrule
        \multirow{6}{*}{\makecell{No \\ Leakage}} 
        & \multirow{2}{*}{MobileNet} & STFT & $0.579 \pm 0.028$ & $0.566 \pm 0.033$ \\
        & & Mel & $0.617 \pm 0.012$ & $0.607 \pm 0.018$ \\
        \cmidrule(lr){2-5}
        & \multirow{2}{*}{ViT} & STFT & $0.620 \pm 0.026$ & $0.608 \pm 0.034$ \\
        & & Mel & $\underline{0.650} \pm \underline{0.027}$ & $\underline{0.640} \pm \underline{0.035}$ \\
        \cmidrule(lr){2-5}
        & \multirow{2}{*}{Swin} & STFT & $0.624 \pm 0.022$ & $0.614 \pm 0.023$ \\
        & & Mel & $\mathbf{0.665} \pm \mathbf{0.012}$ & $\mathbf{0.656} \pm \mathbf{0.017}$ \\
        \midrule
        \multirow{6}{*}{\makecell{Data \\ Leakage}} 
        & \multirow{2}{*}{MobileNet} & STFT & $0.660 \pm 0.002$ & $0.658 \pm 0.003$ \\
        & & Mel & $0.728 \pm 0.003$ & $0.727 \pm 0.002$ \\
        \cmidrule(lr){2-5}
        & \multirow{2}{*}{ViT} & STFT & $\underline{0.827} \pm \underline{0.002}$ & $\underline{0.827} \pm \underline{0.002}$ \\
        & & Mel & $\mathbf{0.837} \pm \mathbf{0.004}$ & $\mathbf{0.837} \pm \mathbf{0.004}$ \\
        \cmidrule(lr){2-5}
        & \multirow{2}{*}{Swin} & STFT & $0.778 \pm 0.002$ & $0.777 \pm 0.002$ \\
        & & Mel & $0.826 \pm 0.004$ & $0.826 \pm 0.004$ \\
        \bottomrule
    \end{tabular}
    \label{table:results}
    \vspace{-0.5em}
\end{table}

\section{Metadata Analysis}
\label{sec:metadata}
There has been little research in how vessel metadata (e.g., speed, distance, size) is correlated to classification accuracy in UATR. In this section, we utilize UniqueShip to explore relationships between classification accuracy of our previously-discussed baseline models and the captured vessel metadata. This will lend intuition to the overall UATR problem and which capture conditions yield reliable classification.

\subsection{Metadata Analysis of Classified Samples:}
To determine the impact of vessel metadata on model performance, we analyze the classification output from our best performing model (Swin with Mel spectrograms in Table \ref{table:results}). Pooling the held-out predictions (validation and test) from all five MMSI-disjoint folds of the ``UniqueShip (5 Class - Balanced)'' split yields 122,878 vessel audio samples. 


We train a secondary classifier on the audio sample's metadata (not the audio) to predict whether the Swin model classified a sample correctly. For this task we utilize CatBoost \cite{catboost}, a gradient-boosting algorithm which handles categorical (MMSI) and continuous (distance, speed) metadata jointly. Predicting classification accuracy above the majority-class baseline indicates that the \textit{metadata inherently encodes the classification difficulty of the vessels}.

%


To disentangle vessel-level from sample-level effects, we adopt a Mundlak formulation \cite{mundlak} based on the ship MMSIs; that is, each metadata variable is split into its per-MMSI mean and the sample's deviation from that mean, separating fixed vessel characteristics (mean, vessel-specific characteristics) from varying capture conditions.


\subsection{Predictive Accuracy Results}
We evaluate the CatBoost model using 5-fold cross-validation on our available samples. Unlike the baseline classification model training, we intentionally permit MMSI leakage in the folds in order to study the influence of MMSIs and the necessity of removing data leakage. Results are depicted in Table \ref{tab:metadata:catboost}. The target baseline accuracy is 0.617, that is, the model blindly guessing that the sample is labeled ``correctly'' will be 61.7\% accurate. This is lower than the reported 66.5\% accuracy in Table \ref{table:results} since we leave out background samples for this analysis. Across the rows, continuous metadata features are incorporated, while the columns represent different treatments of MMSI in the analysis.

\textbf{Impact of Vessel MMSI:} 
Including MMSI as a metadata label (first column) boosts predictive accuracy to approximately 0.768, which marginally increases to 0.802 with additional metadata. This indicates that the model prioritizes MMSI over physical vessel parameters and therefore \textit{individual vessels possess distinct, recognizable signatures that dictate primary classification performance.} This affirms our data leakage hypothesis discussed in the previous sections.

\textbf{Limitations of Distance Metrics:}
Excluding MMSI as a metadata label (second column) makes the accuracy rely heavily on the available metadata, with more metadata producing better prediction accuracy. Notably, using ``Distance'' alone provides minimal improvement over the 0.617 baseline. This reveals that \textit{distance alone is an insufficient metric for determining ship classification viability}, at least within the 2km inclusion radius used in this study. However, incorporating the full metadata suite recovers accuracy to 0.766, approaching the performance of the MMSI-inclusive model.

\textbf{Impact of Ship Metadata:}
Although excluding MMSI initially degrades performance, applying the Mundlak formulation (third column) recovers much of the predictive accuracy. The reason for this improvement is Mundlak disentangles the inter-MMSI metadata. We verified that this advantage persists under strictly non-leaky conditions, confirming the model is not merely using group-level means to implicitly leak MMSI. While the Mundlak formulation requires vessel-grouped data that may be unavailable during real-world deployment, this finding underscores the \textit{intrinsic value of comprehensive physical metadata for building robust, generalizable baseline models}, since the metadata heavily influence classification difficulty.



\begin{figure*}
  \centering
   \includegraphics[width=1\linewidth]{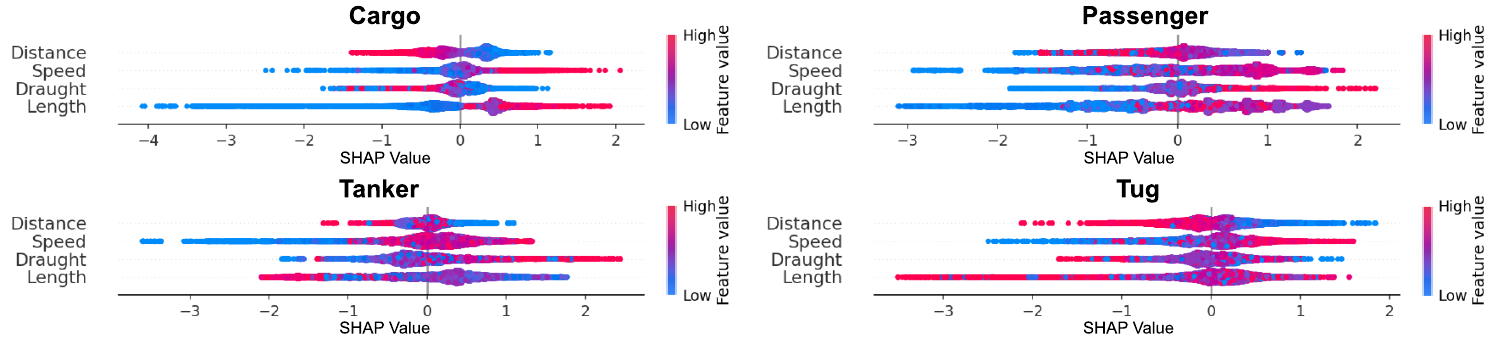}
   \caption{SHAP beeswarms of the four main continuous metadata variables per ship class. A negative SHAP value means that the metadata feature reduces the primary model's classification accuracy, while a positive value relates to improved classification accuracy of the sample. }
   \label{fig:metadata:shap_bee}
   \vspace{-1em}
\end{figure*}

\begin{table}[t]
    \centering
    \footnotesize 
    \setlength{\tabcolsep}{4pt} 
    \caption{Accuracy for predicting correct ship classification based only on metadata (baseline = 0.617).}
    \label{tab:cv_accuracy}
    \begin{tabular}{l c c c}
        \toprule
        & \multicolumn{3}{c}{\textbf{Run Configuration}} \\
        \cmidrule(lr){2-4}
        \textbf{Metadata Features} & \textbf{MMSI} & \textbf{No MMSI} & \makecell{\textbf{No MMSI} \\ \textbf{Mundlak}} \\
        \midrule
        Draught                         & 0.768 & 0.618 & 0.675 \\
        Distance                        & 0.783 & 0.618 & 0.675 \\
        Speed                           & 0.784 & 0.627 & 0.706 \\
        Length                          & 0.763 & 0.678 & 0.690 \\
        Length + Speed                  & 0.785 & 0.725 & 0.746 \\
        Length + Speed + Dist           & 0.797 & 0.730 & 0.758 \\
        Length + Speed + Dist + Draught & 0.800 & 0.750 & 0.768 \\
        All                             & 0.802 & 0.766 & 0.778 \\
        \bottomrule
    \end{tabular}
    \label{tab:metadata:catboost}
    \vspace{-1.5em}
\end{table}

\subsection{SHAP Values}
We can further investigate the CatBoost models on how individual metadata features are related to the predictive performance. SHapley Additive exPlanation (SHAP) measures how much individual features contribute to the prediction outcome, such as whether faster vessels improve or degrade classification accuracy \cite{shapley}. A high SHAP value for a feature relates to ships being correctly classified, while a low SHAP value relates to a ship more likely to be incorrectly classified.

Figure \ref{fig:metadata:shap_bee} depicts SHAP values for each ship type of the when using the top 4 vessel metadata together: distance, speed, draught, and length. Table \ref{tab:metadata:feature_importance} depicts the corresponding average magnitude of the SHAP values in Figure \ref{fig:metadata:shap_bee}, which roughly correspond to how important the features are to overall ship classification. Based on these results, we can conclude that the length and size of the ship typically are the most important metadata for ship classification, followed by the speed, distance, and draught. There also are some intra-class variations, such that bigger length negatively impacts Tanker and Tug classification accuracy but improves Passenger and Cargo accuracy. Overall, this analysis indicates that there is a clear, systematic reliance of ship classification accuracy based on these metadata features, and they can be used to gain intuition on the classification task.


\begin{table}[t]
\centering
\caption{Mean $|$SHAP$|$ Values by Ship Type and Metadata Label}
\label{tab:metadata:feature_importance}
\begin{tabular}{lcccc}
\toprule
\textbf{Ship Type} & \textbf{Length} & \textbf{Speed} & \textbf{Distance} & \textbf{Draught} \\
\midrule
Cargo & \textbf{0.52} & 0.22 & \underline{0.32} & 0.17 \\
Passengership & \textbf{0.76} & \underline{0.67} & 0.30 & 0.30 \\
Tanker & \underline{0.60} & \textbf{0.42} & 0.17 & 0.38 \\
Tug & \textbf{0.36} & \textbf{0.36} & \underline{0.31} & 0.26 \\
\bottomrule
\end{tabular}
\vspace{-1.5em}
\end{table}

\section{Conclusion}


We present \textit{UniqueShip}, a large, labeled dataset for UATR of ships based on the Ocean Networks Canada (ONC) repository, spanning 3,437 hours of audio from 4,218 unique vessels with splits constructed to mitigate data leakage. We train several lightweight UATR baseline models, with our strongest performer (Swin with Mel spectrograms) achieving 0.665 accuracy. We evaluate the effect of data leakage on previous UATR datasets (Deepship \cite{irfan2021deepship} and VTUAD \cite{domingos2022vtuad}) and demonstrate that it inflates accuracy by 10--48 percentage points, a gap that does not translate to real world deployments.

Two results indicate how future UATR datasets should be built. First, our ablations demonstrate that doubling the number of unique vessels gains 2.4--2.6 accuracy points while doubling total audio duration gains only 0.8--1.3. Per doubling, vessel diversity is therefore worth two to three times as much as audio quantity, and gathering unique vessel sources is of greater importance than repeat audio from vessels already captured. Second, our metadata analysis shows that classification difficulty is encoded in vessel properties. Using metadata alone, our gradient-boosting model predicts correct classification with 77.8\% accuracy, a 16 percentage point improvement over the 61.7\% baseline. This analysis also depicts that vessel identity is the most informative feature when splits share MMSIs, further corroborating our data leakage hypothesis. Distance to the hydrophone, however, contributes very little on its own. Together these suggest that curation effort is better spent on breadth of unique vessels than on additional hours, and that vessel characteristics influence sample classification difficulty more than distance alone. We publish the dataset, code, and easy-to-download splits at \datasetlink.

\textbf{Dataset Limitations:} UniqueShip omits audio in which two or more vessels are within the inclusion radius, so models trained do not address the multi-ship cases that occur in real deployments. We apply minimal denoising, which permits some empty or mislabeled samples; we refer to \cite{domingos2022vtuad} for preprocessing options. Our metadata analysis inherits correlations present in the AIS record which may optimistically bias the results in Table \ref{tab:metadata:catboost}, though we expect the effect to be small. Passenger ships have the fewest unique MMSIs (109), making their per-class metadata results the most susceptible to bias. Finally, our published splits are segmented to 5s samples for convenience; full-length recordings are available through the codebase.

\textbf{Future Work:} The 2km inclusion radius used here and elsewhere is a heuristic and fairly unexplored \cite{fischer2024upad}. Our metadata results indicate that a curation criterion combining distance with vessel characteristics may collect more classifiable samples, but identifying the right combination requires a dedicated study. Section \ref{sec:metadata} also suggests the inverse task of our metadata analysis: regressing vessel metadata directly from the audio, which UniqueShip enables but which we have not attempted. A third direction concerns leakage itself. We group ambient recordings by day but vessel recordings only by MMSI; enforcing strict temporal separation, so that each split draws from disjoint days, may reveal residual leakage that MMSI grouping alone does not eliminate. Our repository contains additional splits and the full 12-class corpus. 





\section{Acknowledgment}

We thank the authors of the prior ONC-based datasets \cite{irfan2021deepship, domingos2022vtuad, li2024oceanship}, especially \cite{domingos2022vtuad} whose code \cite{cesar_lucascesarfdonc_dataset_2026} was helpful to this work.

This material is based on work supported by the Defense Advanced Research Projects Agency (DARPA) under Maritime Acoustic Recognition \& Identification with Novel Algorithms (MARINA), contract number W912CG-24-C-0020. 

\textbf{Distribution Statement A.} Approved for public release: distribution is unlimited.

\section{Disclaimer}

\noindent The views, opinions, and/or findings expressed are those of the author(s) and should not be interpreted as representing the official views or policies of the Department of Defense or the United States Government.

\bibliographystyle{IEEEtran}
\bibliography{references.bib}

\clearpage

\onecolumn
\section{Supplementary/Additional Figures}
\subsection{Codebase Flow}

\begin{figure}[h]
  \centering
   \includegraphics[width=1\linewidth]{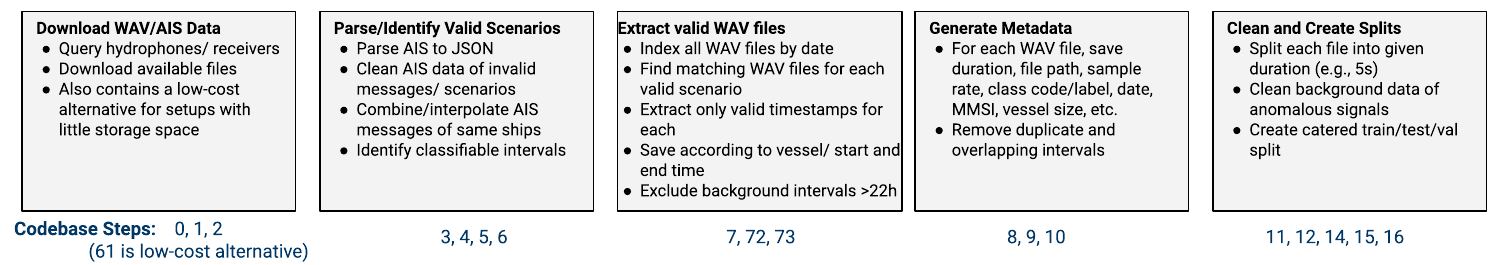}
   \caption{Workflow of codebase repository.}
   \label{supp:fig:coderepo}
\end{figure}

Figure \ref{supp:fig:coderepo} depicts the overall workflow of the codebase published at the website (\datasetlink) which generates the dataset and the resulting splits.

There are five main steps: the first downloads all the WAV files and AIS data needed. This is a very space-intensive operations, as the entire wav and AIS repository is several TBs. There is an optional method supported that instead just queries the specific files for the identified scenarios, but this can be slower if making multiple datasets with changing radii. The second step parses and identifies valid scenarios in which 1) one vessel is within 2km (inclusion radius) of a hydrophone and no others within 4km, and 2) no vessels are within 8km (exclusion radius) of a hydrophone (i.e., background data). This is all focused on the AIS data, and involves linear interpolation in between AIS pings to interpolate the vessel positions. The third step extracts the valid WAV files according to the valid intervals for the specific inclusion and exclusion radii. We also remove background intervals greater than 22h since we found these to be mislabeled and to actually contain many ships. The fourth step is to extract the important metadata from the AIS signals for each audio file and clean up the intervals to remove any duplicates or overlaps. Finally, the last step partitions the audio into 5s segments and creates the different splits. This is where the ``leakproof'' criteria are enforced.

\subsection{Temporal Distribution of Data}
\begin{figure}[!h]
  \centering
   \includegraphics[width=.6\linewidth]{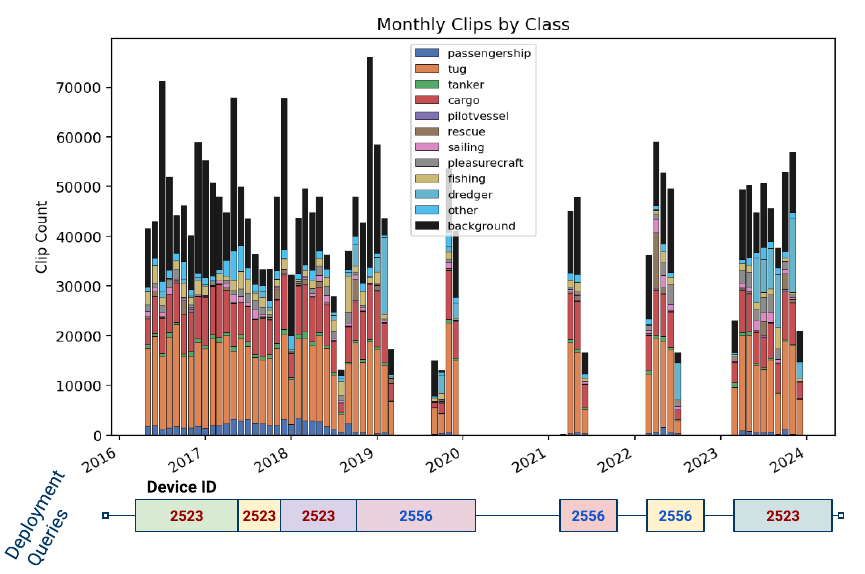}
   \caption{Temporal distribution of deployments \cite{ONC_2556_2022, ONC_2556_2021, ONC_2556_2018, ONC_2523_2023, ONC_2523_2017b, ONC_2523_2017a, ONC_2523_2016} 5s audio samples.}
   \label{supp:fig:temporal}
\end{figure}

Figure \ref{supp:fig:temporal} depicts the total 5 second clips generated per month across all the deployments. The dataset uses two hydrophones (IDs 2523 and 2556) and seven deployments \cite{ONC_2556_2022, ONC_2556_2021, ONC_2556_2018, ONC_2523_2023, ONC_2523_2017b, ONC_2523_2017a, ONC_2523_2016} as depicted in the bottom row of the figure. There are some temporal gaps in the stream where either the data is not public or not collected.

\subsection{Additional Result: Confusion Matrix}
\begin{figure}[htbp]
  \centering
   \includegraphics[width=.5\linewidth]{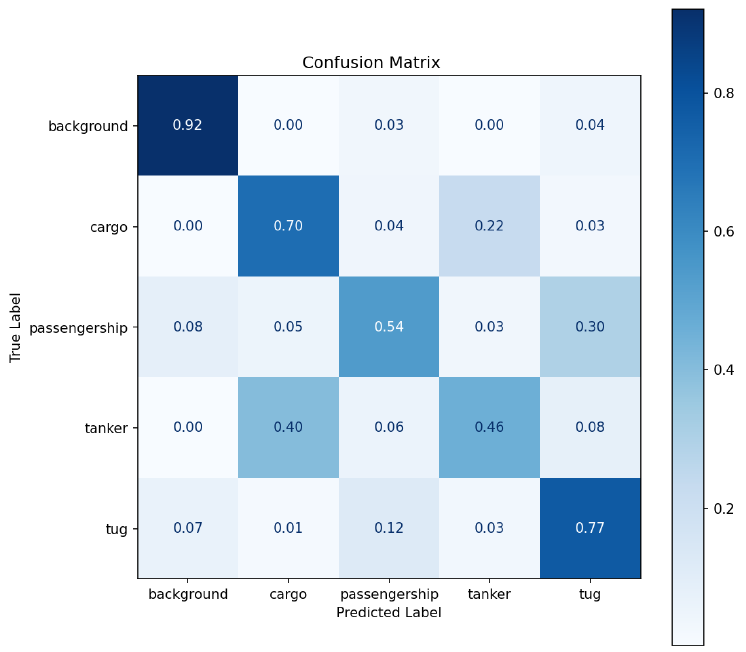}
   \caption{Confusion matrix across all samples in the ``Main 5 - Bal.'' split with the top performing model in Table \ref{table:results} (Swin model with Mel spectrograms). The results are combined across the 5 different seeds.}
   \label{fig:supp:confusion}
\end{figure}

Figure \ref{fig:supp:confusion} displays the confusion matrix for the ``5 Class - Balanced'' split with the top performing model in Table \ref{table:results} (Swin model with Mel spectrograms). From this, we see that background is fairly easy to identify compared to the other classes. This makes sense since it is mostly the absence of a signal, instead of classifying between different signals. Cargo and Tanker have the most overlap, with tanker having the worst score overall. Future work could focus more on the relationship between these two classes and why they are often mixed together, perhaps due to similar vessel characteristics.

\subsection{Results of CatBoost on No-Leakage Splits}

Table \ref{supp:tab:catboost_noleak} replicates the CatBoost analysis presented in Table \ref{tab:metadata:catboost} of the main text but enforces a non-leaky training/test split for training the CatBoost models (i.e., no MMSIs shared between the splits). Note that the primary acoustic classification model (Swin) was already evaluated without data leakage, as this analysis specifically addresses the training environment of the secondary CatBoost model to validate our interpretation of the metadata features.

Two things to note from these non-leaky results

\begin{itemize}
    \item The first is that when the MMSI is included in the non-leaky setup (first column), the model's accuracy drops essentially baseline to baseline (i.e., 0.617) across all metadata configurations. This is much different than the previous results with the leaky setup in Table \ref{tab:metadata:catboost} of the main text in which it performs the best over all the runs. Because the training and testing sets no longer share MMSIs yet MMSIs still contain significant information, the model's tendency to memorize individual MMSIs renders it incapable of predicting classification success on unseen ships. \textbf{This reinforces our earlier conclusion that models will overfit to MMSIs if permitted,} resulting in less generalization power. 
    \item The second conclusion is based on the ``No MMSI Mundlak'' configuration (third column), and that it still has significant improvement over without using Mundlak formulation (second column). This confirms our hypothesis that the Mundlak formulation is \textbf{not merely imitating MMSIs but instead better capturing metadata variations}. By isolating each MMSI's average profile, the Mundlak formulation captures generalizable and intrinsic characteristics.
\end{itemize}

\begin{table}[htbp]
    \centering
    \footnotesize 
    \setlength{\tabcolsep}{4pt} 
    \caption{Cross-Validation Accuracy (No leak) by Metadata and Run Configuration. The baseline accuracy for the model is 0.617.}
    \label{tab:cv_accuracy}
    \begin{tabular}{l c c c}
        \toprule
        & \multicolumn{3}{c}{\textbf{Run Configuration}} \\
        \cmidrule(lr){2-4}
        \textbf{Metadata Features} & \textbf{MMSI} & \textbf{No MMSI} & \makecell{\textbf{No MMSI} \\ \textbf{Mundlak}} \\
        \midrule
        Draught                         & 0.617 & 0.618 & 0.675 \\
        Distance                        & 0.617 & 0.617 & 0.678 \\
        Speed                           & 0.617 & 0.627 & 0.706 \\
        Length                          & 0.617 & 0.678 & 0.690 \\
        Length + Speed                  & 0.617 & 0.725 & 0.746 \\
        Length + Speed + Dist           & 0.619 & 0.731 & 0.759 \\
        Length + Speed + Dist + Draught & 0.619 & 0.752 & 0.766 \\
        All                             & 0.623 & 0.766 & 0.778 \\
        \bottomrule
    \end{tabular}
    \label{supp:tab:catboost_noleak}
\end{table}


\subsection{Additional Metadata Analysis for ``Main 5'' Split}

\subsubsection{Vessel Metadata Correlations}
\begin{figure}[h]
  \centering
   \includegraphics[width=.5\linewidth]{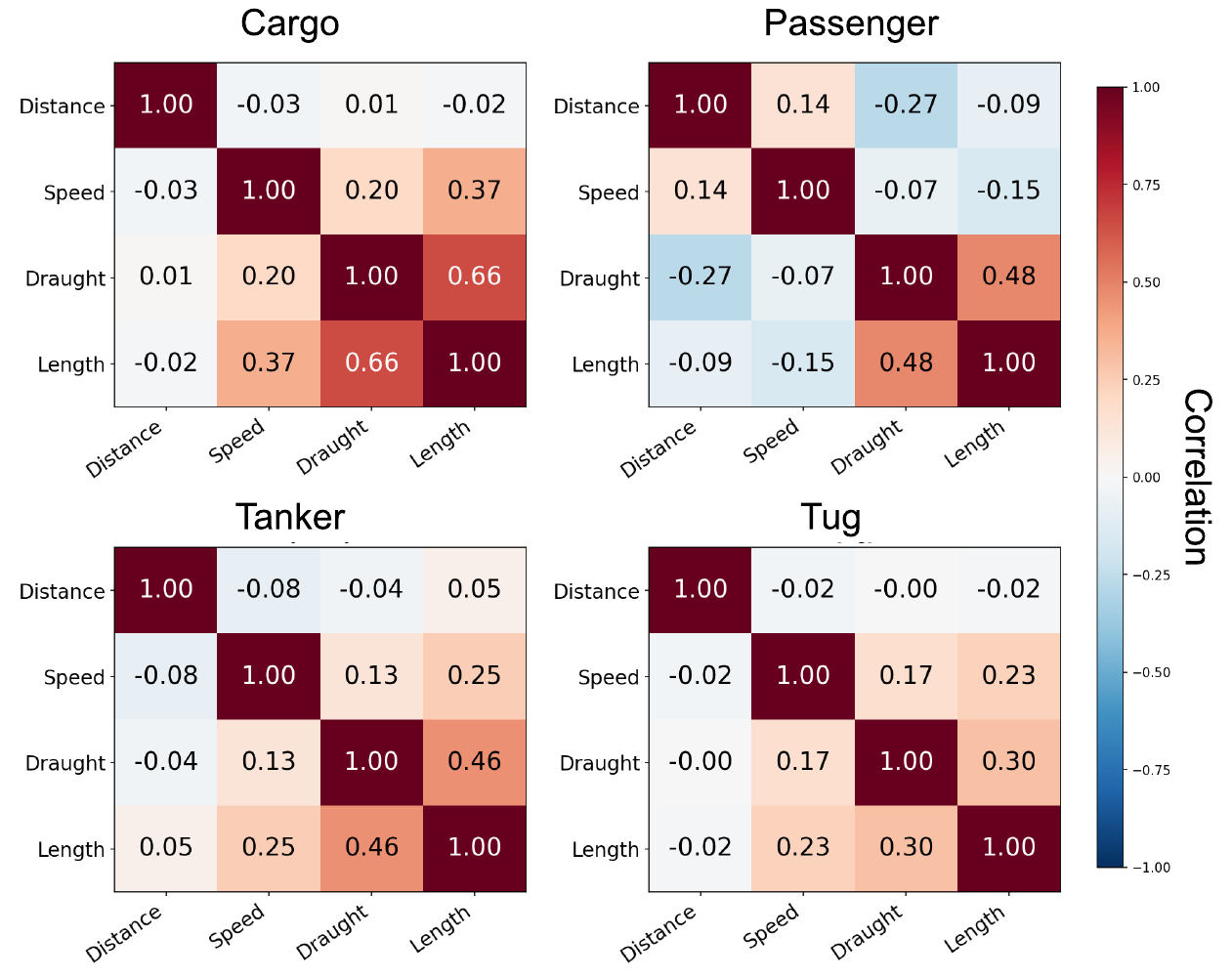}
   \caption{Pearson correlation coefficient of ship metadata based on the ship class, where a high absolute value (max 1.0) indicates strong correlation.}
   \label{fig:metadata:corr}
\end{figure}

To investigate the usefulness of the vessel metadata, we plot the correlations between them to see if dependencies arise, as highly-correlated metadata are often redundant. Fig \ref{fig:metadata:corr} depicts correlations between the several key vessel metadata for each vessel class. Some of the correlations are likely intrinsic, such that the draught is reliant on ship length and speed. However, while we expect distance to be uncorrelated with any quantity, it has some correlations, especially with passenger ships. Same for the correlations in speed and length. These are likely some biases in our dataset which could ultimately affect our analysis, such that passenger ships collected far away from hydrophones also typically are faster or have less draught. Outside of passenger ships, these unexpected correlations remain fairly small, besides how ship length is correlated with speed for most ship types.

\subsubsection{Additional Figures}
\begin{figure}[htbp]
  \centering
   \includegraphics[width=1\linewidth]{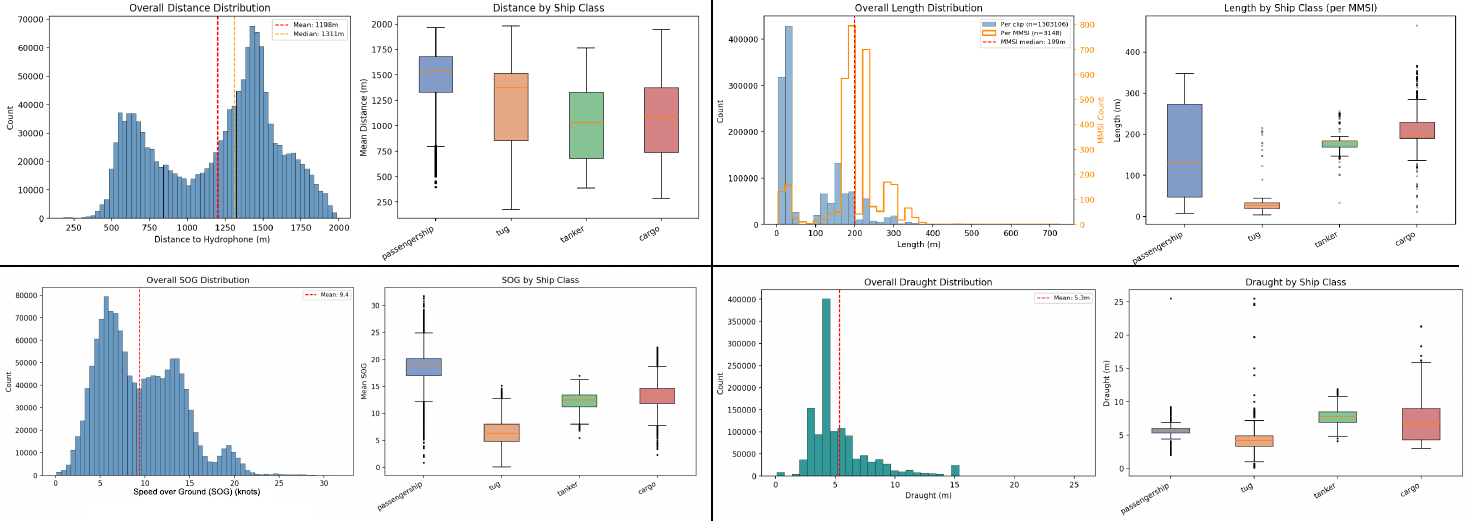}
   \caption{Distribution of 4 vessel characteristic metadata features (distance, speed, draught, and length) across the main four vessel classes.}
   \label{supp:fig:metadata_distributions}
\end{figure}

\begin{figure}[htbp]
  \centering
   \includegraphics[width=.5\linewidth]{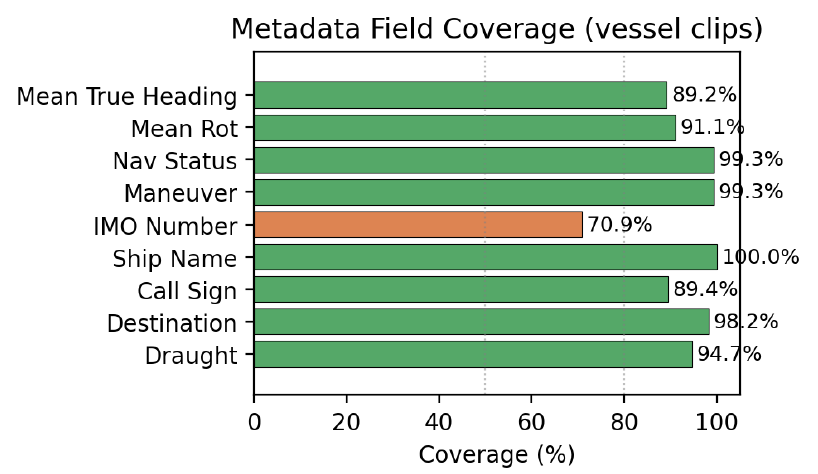}
   \caption{Metadata fields (minus CTD information) which do not have full coverage in dataset. Note that while ``ship name'' records 100\% in the figure, it is missing very few labels (near 99\%). CTD is not included in this table, and the current version of the dataset does not fully extract CTD over all deployments.}
   \label{supp:fig:metadata_fill}
\end{figure}

Figure \ref{supp:fig:metadata_distributions} displays the distribution of some of the metadata features across the main four vessel classes. Figure \ref{supp:fig:metadata_fill} displays the percent of fields populated across the main split. Fields which have 100\% coverage are not included in this figure.

\end{document}